# Shape-Preserving Accelerating Electromagnetic Wavepackets in Curved Space

Rivka Bekenstein, Jonathan Nemirovsky, Ido Kaminer and Mordechai Segev

Physics Department and Solid State Institute, Technion, Haifa 32000, Israel

**Abstract**

We present shape-preserving spatially accelerating electromagnetic wavepackets in curved space: wavepackets propagating along non-geodesic trajectories while recovering their structure periodically. These wavepackets are solutions to the paraxial and non-paraxial wave equation in curved space. We analyze the dynamics of such beams propagating on surfaces of revolution, and find solutions that carry finite power. These solutions propagate along a variety of non-geodesic trajectories, reflecting the interplay between the curvature of space and interference effects, with their intensity profile becoming narrower (or broader) in a scaled self-similar fashion  Finally, we extend this concept to nonlinear accelerating beams in curved space supported by the Kerr nonlinearity. Our study concentrates on optical settings, but the underlying concepts directly relate to General Relativity.

The complex dynamics of particles and of electromagnetic (EM) waves in curved space-time is still inaccessible to laboratory experiments. However, numerous physical systems have been suggested to demonstrate analogies of General Relativity phenomena, ranging from sound and gravity waves in flowing fluids [1–3], to Bose-Einstein [4–6] and optical systems, which have had a major success in demonstrating such phenomena [7–13]. For example, metamaterials enabled creating analogies to black holes, by engineering the (EM) properties of the material through which light is propagating [8–10]. Another example is using a moving dielectric medium that acts as an effective gravitational field on the light [12]. This idea was demonstrated experimentally by employing ultrashort pulses in an optical fiber to create an artificial event horizon [13]. Another route for such studies is to create curved space by engineering the geometry of the space itself. This idea, suggested in 1981 [14], started by exploring the dynamics of a free quantum particle constrained by an external potential to evolve within a thin sheet. More than 25 years later, these ideas were carried over to EM waves [15], where pioneering experiments studied light propagating in a thin film waveguide attached to the curved surface area of a three-dimensional (3D) body [16]. However, thus far, in all of these experiments and theoretical studies on General Relativity concepts with EM waves - the wavepackets were propagating on geodesic trajectories, which are naturally the shortest path, analogous to straight lines in flat geometry. But, do wavepackets propagating in curved space have to follow special geodesic paths, or can they exhibit other trajectories that are not predicted by the geodesic equation?

Here, we show that wavepackets can exhibit periodically-shape-invariant spatially-accelerating dynamics in curved space, propagating in non-geodesic trajectories that reflect interplay between the curvature of space and interference effects arising from initial conditions. We study these beams in the linear and nonlinear, paraxial and nonparaxial regimes, and unravel a variety of new intriguing properties that are nonexistent in flat space. This study paves the way to accelerating beams experiments in curved space to study basic concepts of General Relativity, where the entire dynamics in non-geodesic.

Before proceeding, we briefly recall the ideas underlying accelerating wavepackets. They were first revealed in 1979 as a unique solution to the Schrodinger equation: a propagation-invariant wavepacket shaped as an Airy function that accelerates in time [17]. Almost 30 years later, the concept of accelerating wavepackets was introduced into electromagnetism, demonstrating Airy beams that are spatially accelerating within the paraxial approximation [18,19]. Following the work of [18,19], accelerating wavepackets have drawn extensive interest and initiated many new ideas, such as accelerating ultrashort pulses and light bullets [20–22], two-dimensional (2D) accelerating beams [23], accelerating beams following arbitrary convex acceleration trajectories [24,25], and accelerating beams in nonlinear media [26–29]. These were followed by many applications such as manipulating micro-particles [30], self-bending plasma channels [31] and accelerating electron beams [32]. For some time, shape-preserving accelerating wavepackets were believed to exist strictly within the domain of the Schrodinger-type paraxial wave equation [17–19]. However, last year we presented accelerating shape-invariant wavepackets that are exact solutions of Maxwell's equations [33]. Experimental demonstrations of such beams followed soon thereafter [34–36], along with further theory and experiments demonstrating additional families of non-paraxial accelerating beams [37–39]. Thus far, however, accelerating wavepackets remained strictly within the realm of flat space.

Since the dynamics of EM waves in curved space is significantly different from that in flat space, a natural question to ask is whether accelerating wavepackets can at all exist in curved space, and if they do how do their features differ from those in flat space. In other words, are there wavepackets that travel along non-geodesic trajectories in free-space without contradicting the basic concepts of General Relativity?

Consider EM waves restricted to exist in 2D curved surface. This can be achieved by covering the surface area of a 3D shape (a sphere, for example) with a thin homogenous layer of a material with a higher refractive index. Such a layer acts as a waveguide, keeping the light confined inside it due to total internal reflection (Fig. 1). The dynamics of EM fields in curved space can be described by the 3D Maxwell equations in general coordinates [40]:

$$\frac{1}{2\sqrt{g}}\varepsilon^{\alpha\beta\gamma}\left(\frac{\partial E_\gamma}{\partial x^\beta}-\frac{\partial E_\beta}{\partial x^\gamma}\right)+\frac{1}{c}\frac{\partial B^a}{\partial t}=0 \qquad \frac{1}{\sqrt{g}}\frac{\partial}{\partial x^a}\left(\sqrt{g}B^a\right)=0$$
$$\frac{1}{2\sqrt{g}}\varepsilon^{\alpha\beta\gamma}\left(\frac{\partial H_\gamma}{\partial x^\beta}-\frac{\partial H_\beta}{\partial x^\gamma}\right)-\frac{1}{c}\frac{\partial D^a}{\partial t}=0 \qquad \frac{1}{\sqrt{g}}\frac{\partial}{\partial x^a}\left(\sqrt{g}D^a\right)=0 \tag{1}$$

Here, $g$ is the time-independent spatial metric determinant where $ds^2 = g_{\alpha\beta}dx^\alpha dx^\beta$ (the spatial indices $\alpha, \beta, \gamma$ run from 1 to 3), $\varepsilon^{\alpha\beta\gamma}$ is the antisymmetric Levi-Civita tensor, $E^\alpha, H^\alpha, D^\alpha, B^\alpha$ are three vectors. The wave equation for the electric field is derived from Eqs. (1) [15]:

$$-\frac{1}{\sqrt{g}}\partial_\beta \sqrt{g} g^{\beta\gamma}\partial_\gamma E^\alpha + \frac{1}{\sqrt{g}}\partial_\beta \sqrt{g} g^{\alpha\gamma}\partial_\gamma E^\beta + \frac{1}{c^2}\frac{\partial^2 E^\alpha}{\partial t^2} = -\frac{1}{c^2}\frac{\partial^2 P^\alpha}{\partial t^2} \quad (2)$$

where the polarization $P^\alpha \equiv D^\alpha - E^\alpha$ can generally be nonlinear in the electric field. Notice that the second term does not appear in homogenous flat space: it arises strictly due to the curved space geometry.

We are interested in the evolution of the electric field in a general surface of revolution. First, we introduce the metric of such a surface. These surfaces are parameterized by $\vec{s}(u,v) = (\alpha(u)\cos(v), \alpha(u)\sin(v), \beta(u))$, where $v = [-\pi, \pi]$ is the angle of rotation and $-\infty < u < \infty$ is a general parameterization of the surface along its axis of the revolution. Every point in 3D space $(\vec{r})$ can be described by the two coordinates on the curved surface $(u,v)$ and a third coordinate $(h)$ normal to the surface at every point: $\vec{r}(u,v,h) = \vec{s}(u,v) + h\vec{N}(u,v)$, where $\vec{N}(u,v)$ is the unit vector normal to the surface (Fig 1). We transform to a new set of coordinates: $z = \int_0^z \sqrt{\alpha'^2(u) + \beta'^2(u)}du$ and $x = R_0 v$, where $R_0$ is defined by the radius of the surface at $z = 0$, $R_0 = \alpha(0)$, and $x$ has units of length in the transverse direction at $z = 0$. The metric takes the form $dl^2 = dz^2 + [\alpha^2(u(z))/R_0^2]dx^2 \triangleq dz^2 + \gamma dx^2$ where $\gamma$ is defined as the dimensionless 2D metric determinant. As in [15], we decouple the wave equation for the different polarizations [14]. This can be done for surfaces that have small enough Gaussian and mean curvatures, and that their mean curvature varies on scales large compared with the wavelength. For example, for a wavelength in the visible range the radius of curvature of such surface has to be of the order of millimeters, a regime in which practically all macroscopic optical components exist (the exceptions are microlenses, microcavities, single-mode fibers, etc.). We are interested in waves propagating in the $z$-direction. The simplest cases are the TE modes, for which the electric field has no $z$-component, hence they are $x$-polarized, in the form $\vec{E}(z,x,h) = (0, \phi(z,x)\xi(h), 0)$, which yields

$$\frac{1}{\gamma}\partial_x^2\phi + \partial_z^2\phi + \frac{\sqrt{\gamma_z}}{\sqrt{\gamma}}\partial_z\phi + q^2\phi + V_{NL}(\phi)\phi = 0 \tag{3.1}$$

$$-\partial_h^2\xi - k_0^2 n_0^2\xi = -q^2\xi \tag{3.2}$$

Here $n_0$ is the refractive index in the surface layer, $k_0$ is the vacuum wavenumber and $q$ has units of $[1/m]$.

The boundary conditions here yield some unexpected implications. Naturally, beams propagating on surfaces of revolution must fulfill periodic boundary conditions for every $z$. Thus, we first find solutions in an infinite space and then use their superpositions to construct solutions satisfying periodic boundary conditions. This methodology serves as a powerful tool to find the solutions in the linear regime $(V_{NL}(\phi) = 0)$ where superposition holds. To do that, we use the universal covering space: a covering map of an infinite 1D space mapped to a ring on the surface (points having the same $z$). Each point on the surface is an image of an infinite number of points located in the universal covering space. We use the covering map to construct solutions as follows

$$\phi_p(z, x_p) = \sum_{m=-\infty}^{\infty} \phi(z, x + 2\pi m R_0) \tag{4}$$

$\phi_p$ is a solution of (3.1) satisfying periodic boundary conditions, where $x_p, x \in [0, 2\pi]$. Equation (4) reflects the fact that Eq. (3.1) in linear in $\phi$ (that is, when $V_{NL}(\phi) = 0$).

First, we focus on the paraxial regime, and derive the equation for the slowly varying amplitude $\psi(z,x)$, assuming that $|\partial^2\psi/\partial z^2| \ll |2q\,\partial\psi/\partial z|$. We use the Ansatz $\phi(z,x) = \frac{1}{\sqrt{\gamma}}\psi(z,x)e^{iqz}$, where the field amplitude $\phi(z,x)$ varies with the algebraic factor $\sqrt{\gamma}$,

for the power to be conserved. This yields the paraxial equation for a general surface of revolution:

$$2iq\partial_z\psi = -\frac{1}{\gamma}\partial_x^2\psi - V_{eff}(z)\psi - V_{NL}(\psi)\psi \qquad (5)$$

where the effective one-dimensional potential depends on the determinant of the surface $V_{eff}(z) = \left(\sqrt{\gamma}/(1/\sqrt{\gamma})_z\right)_z$. Clearly, the paraxial equation describing the propagation of EM waves within surfaces of revolution involves more complex evolution than the propagation of an optical beam in flat space. First, the surface curvature acts as a $z$-dependent one-dimensional potential even for homogeneous surfaces. Second, the spatial frequencies vary during propagation, in analogy to the redshift and blueshift occurring in curved space-time. Consequently, the shapes of the eigenmodes describing the waves propagating in such a surface evolve when the curvature of space varies during propagation.

We seek an accelerating solution to Eq. (5), namely a solution that is propagation-invariant in the accelerating frame of reference. Such solutions should satisfy $|\psi(0,x)| = |\psi(z, x-f(z))|$ meaning that the beam would propagate along the curve $x = f(z)$ while maintaining its intensity structure. We want to transform Eq. (5) to the paraxial equation in flat space and use the known solution of the accelerating Airy beam. To do that, we first cancel the effective potential using a gauge transformation $\tilde{\psi} = \psi e^{-\frac{i}{2q}\int_0^z V_{eff}(z')dz'}$. Then, we use a transformation of coordinates of the form: $\tilde{z} = \int_0^z \frac{1}{\gamma(z')}dz'$, and find the accelerating beam in curved space to be:

$$\psi(z,x) = Airy(x - f(z))\exp\left(i\left(\frac{a}{2q}\int_0^z \frac{1}{\gamma}dz'\right)(x - f(z)) + \frac{i}{2q}\int_0^z V_{eff}(z')dz'\right) \qquad (6)$$

Where $a$ is a constant with units $[a] = 1/m^3$. The expression for the trajectory $(f(z))$ of the Airy beam in curved space is given by:

$$f(z) = a\left(\frac{1}{2q}\int_0^z \frac{1}{\gamma}dz'\right)^2 \qquad (7)$$

Equation (7) defines acceleration trajectories that depend on the metric determinant. Consequently, the acceleration trajectory is different for every surface of revolution, and can even become non-convex in $x$, as shown in Fig. 1. Notice that, generally, the accelerating solution of Eq. (5) is not shape-preserving because $|\psi|^2$ varies with $z$. However, it is self-similar and can become narrower or broader during propagation, according to the geometry of the specific surface [41].

To understand the origin of the non-geodesic trajectory, we introduce a particle model to describe the trajectory of the main lobe of the accelerating beam. We account for the interference effect through an inhomogeneous term in the geodesic equation:

$$\frac{d^2x}{d\lambda^2} + \frac{\gamma_z}{\gamma}\frac{dx}{d\lambda}\frac{dz}{d\lambda} = \frac{\tilde{F}}{\sqrt{\gamma}} \tag{8}$$

where $\lambda$ is an affine parameter which, in this case, can be the line element. This equation describes the motion of a particle in a surface of revolution under the influence of a force, where $\tilde{F}$ has the dimensions of force per unit of mass when $\lambda$ is taken to be time. Obviously, $\tilde{F}$ is a "fictitious" force, because no real force is acting here. Constraining the motion of the particle to "paraxial" motion, $|dx/dz| \ll 1$, yields the approximate line element $d\lambda = dz\left(1 + \frac{1}{2}dx^2/dz^2 + O\left(\left(dx^2/dz^2\right)^2\right)\right)$. To first order, Eq. (8) becomes:

$$\frac{d^2x}{dz^2} + \frac{\gamma_z}{\gamma}\frac{dx}{dz} = \frac{\tilde{F}}{\sqrt{\gamma}} \tag{9}$$

Here, the fictitious force, $\tilde{F} = a/k\left(\sqrt{\gamma}\right)^3$, which manifests the interference effect induced by the structure of the wavepacket, and also reflects the dependence on the curvature of the surface of revolution. This is a unique wave phenomenon that a particle model cannot describe. While propagating on a surface of revolution, the arc length in stretched as $d\lambda = \sqrt{\gamma}dx$, hence the spatial frequencies of the beam are stretched with an opposite trend. This changes the fictitious force $\tilde{F}$ by a factor of $\left(\sqrt{\gamma}\right)^3$, as can be seen directly from the cubic phase of the accelerating beam in $k$-space. The solution for $x(z)$ in Eq. (9) is exactly the trajectory of the Airy beam - $f(z)$ from Eq. (7).

The non-parabolic acceleration trajectories in curved space can be understood by examining Eq. (9), which manifests the interplay between the effect of the curvature and the effect of interference. The right hand side comes from the interference effect acting as if an effective potential exerts a "fictitious" force on the wavepacket. In flat space, the second term on the left is zero because $\gamma = const$, and the equation becomes the Newton equation for a particle under a constant force, which yields a parabolic trajectory. The same parabolic trajectory is the trajectory of the Airy beam in flat space. Clearly, the curvature of space has a major effect on the trajectory of the beam, through the "fictitious" force. However, the curvature also gives rise to another term in Eq. (9): the second term, $\gamma_z/\gamma$, which is one of the two non-zero Christoffel symbols.

Thus far, we generalized the paraxial accelerating beam to curved space, and showed the various trajectories possible which are not the natural geodesics of these surfaces, but we did not find the actual solutions as of yet. To do that, we construct a beam propagating on the trajectories defined by Eq. (7) and also fulfills periodic boundary conditions, as necessary for surfaces of revolution. Such solutions are naturally periodic [42] and they are obtained from the Airy solution defined on the universal covering space, using Eq. (4):

$$\psi(z, x_p) = \sum_{m=-\infty}^{m} C_m \exp\left(\frac{ik_m^3}{3a} + ik_m x\right) \exp\left(i\left(\left(\frac{a}{2q}\int_0^z \frac{1}{\gamma} dz'\right)x + \frac{1}{2q}\int_0^z V_{eff}(z')dz'\right)\right) \qquad (10)$$

where $k_m = m/R_0$ and $m$ is an integer. The initial beam (the beam at $z = 0$) is actually an infinite Airy beam that is wrapped on a circular perimeter, over and over again. This solution satisfies Eq. (5), and also the periodic boundary conditions. These conditions can be satisfied only by specific (quantized) values of transverse momentum. Hence, the beam is composed of a discrete set of "spectral functions". To stay within paraxiality, we limit the spatial spectrum from above, by setting $C_m = 0$ for every spatial frequency above the $k_M$ defining the boundary of the paraxial regime. Importantly, the number of these spectral functions comprising the beam is constrained both from below and from above: the lowest transverse wavenumber that can be excited when the beam is launched (at $z = 0$) is $k_1 = 1/R_0$, while the highest $k_M = M/R_0$ occurs for $m = M$. This finite range within which the spatial frequencies of the accelerating beam can exist has immediate physical consequences: such a curved-space accelerating beam carries finite power, because it is constrained to a circular perimeter and constructed from a finite number of spatial frequencies, due to the cut-off. This finding has an important implication: having a finite power, one can now define a center of mass for the accelerating beam. It is important to emphasize that although the self-reconstructing structure of the wavepacket travels along a non-geodesic trajectory, the center of mass travels along a geodesic trajectory as in [17,18]. However, almost all the applications of accelerating beams rely on light-matter interactions, where the important parameter is the local intensity and not the center of mass, e.g., acceleration of particles [30], formation of curved plasma channels [31], laser machining [43], to name a few out of many. For all such applications, what matters is the accelerating main lobe where the intensity is the highest, while the fact that the center of mass is propagating on a straight line is unimportant.

In examining the structure of the curved-space accelerating beam, we notice that it can be different from the Airy beam whose envelope is monotonically decaying. Here, the shape-preserving wavepacket accelerating in curved space can have several parallel beams whose number is set by the initial choice of the spectral components $C_m$.

The accelerating solution in curved space is propagating on the curve defined by Eq. (7) and is periodically shape-invariant: it recreates its exact intensity profile in $x_p$ for discrete $z$-values defined by $\int_0^{z_q} \frac{1}{\gamma} dz' = \frac{2ql}{aR_0}$ where $l$ is an integer. Notice the non-constant spacing between planes of self-reconstruction that depends on the curvature of space. This interesting feature results from the transverse momentum being quantized (rather than continuous). Interestingly, $a$ determines also the curvature of the trajectory: the faster the beam accelerates the faster it recreates itself.

After having presented the paraxial accelerating beams in curved space and their properties, we now proceed to the non-paraxial beams which are solutions of Maxwell's equations on surfaces of revolution. We begin with Eq, (3.1), which describes the linear non-paraxial regime.

We apply transformation of coordinates that simplifies the equation for any surface of revolution. We set: $Z = \int_0^z \frac{1}{\sqrt{\gamma(z')}} dz'$ which yields:

$$\partial_x^2 \phi + \partial_Z^2 \phi + \gamma q^2 \phi = 0 \tag{11}$$

Clearly, the non-paraxial case is more complicated than the paraxial one: Eq. (11) is essentially the Helmholtz equation with a $z$-dependent refractive index. This is an equation that allows back propagation and back reflections. Here, we look only for a forward propagating wavepacket. Since we cannot solve at this point for the most general case, we examine three generic solutions which allow for close-form solutions. The first case of a surface of revolution is a cylinder, where the metric determinant is not $z$-dependent $\gamma(z) = 1$. The solution in the covering space coincides with the form of the solution in flat space, which is described in details in [33]. Using the same method we used for the paraxial beam (Eq. 4), we construct the accelerating wavepacket:

$$\hat{\phi}(Z, x_p) = \sum_{q_n} D_n \exp\left(i\beta q_n + iq\left(x_p \cos(q_n) + Z \sin(q_n)\right)\right) \tag{12}$$

We choose $D_n = 0$ for any $q_n$ that is not between 0 and $\pi$, meaning that we allow only forward-propagating waves (i.e., we assume that the backward-propagating waves are not excited at $z = 0$). This wavepacket is constructed from a discrete set of spatial frequencies that fulfill the periodic boundary conditions: $q_n = \arccos(n/qR_0)$ (see Fig. 2). The spectrum is now limited from above, because at a high enough spatial frequency the propagation constant becomes imaginary and the spectral function becomes evanescent. Here, we are not interested in the evanescent waves, hence we set their initial population to zero ($D_n = 0$ for those modes). This non-paraxial accelerating beam carries finite power. In fact, the solution can support several parallel beams accelerating (bending) in parallel, as in the paraxial case, for a suitable choice of $D_n$. As for the nonparaxial flat-space accelerating beams [33] this nonparaxial curved-space wavepacket is approximately shape-invariant because it is a superposition of only forward propagating waves ($0 < q_n < \pi$). If the counter-propagating waves were to be taken in the superposition, the beam would have been fully shape-preserving. Nevertheless, this wavepacket (Fig. 2) accelerates on a circular trajectory while bending to very large (almost 90°) non-paraxial angles. The beam reconstructs itself in discrete angles in the $x, z$ plane, specifically for $\theta_n = \arccos(n/qR_0)$. We point out, however, that as $R_0$ becomes smaller - there are less propagating modes, until eventually, when $R_0$ becomes smaller than the wavelength of the light, all the excited spatial functions are evanescent. Equation (12) defines a family of solutions for a given trajectory, where every $\beta$ gives a beam with a different structure. Thus, every superposition of such beams (of various values of $\beta$) also forms a periodically shape preserving accelerating beam.

Having solved for the simplest non-paraxial surface of revolution (a cylinder, where the metric is not z-dependent), the natural question to ask is whether a non-paraxial accelerating shape-invariant solution can exist for surfaces with a z-dependent curvature. Finding these kinds of solutions is especially challenging, because they cannot rely on the symmetry between all space coordinates, since this symmetry is inherently broken. Going back to Eq. (3.1), we simplify the equation using: $\phi(z,x) = 1/\gamma^{1/4}\, \zeta(z,x)$, which cancels the term with the first derivative in respect to $z$, yielding:

$$\partial_x^2 \zeta + \gamma \partial_z^2 \zeta + \gamma\left(\frac{1}{4}\left(\left(\sqrt{\gamma}\right)_z\right)^2 \Big/ \gamma - \frac{1}{2}\left(\sqrt{\gamma}\right)_{zz}\Big/\sqrt{\gamma} + q^2\right)\zeta = 0 \tag{13}$$

This equation is a Helmholtz type equation with two differences: (1) there is an additional term that gives a $z$-dependent addition to the effective wavenumber, and (2) the $z$-dependent metric multiplies all the terms except for the derivative with respect to $x$. We want to transform Eq. (13) to a Helmholtz equation with a constant "effective wavenumber". For this cause, we choose two specific surfaces; one with positive curvature and one with negative curvature, that will give a constant effective wavenumber; $\gamma_p = \cos^4(\kappa z), \gamma_n = \cosh^4(\kappa z)$. Equation (13) then simplifies to:

$$\partial_x^2 \zeta / \gamma + \partial_z^2 \zeta + \left(q^2 \pm \kappa^2\right)\zeta = 0 \tag{14}$$

where the $\pm$ sign stands for the positive and negative curvatures, respectfully. Following the same approximation regarding the slow change in curvature on the scale of a wavelength, we assume that $\kappa \ll 1/R_0$. We find the accelerating wavepackets on these surfaces to be:

$$\hat{\phi}(z, x_p) = 1/\left(\gamma_{1,2}\right)^{1/4} \sum_{q_n} D_n \exp\left(i\beta q_n + i\sqrt{q^2 \pm \kappa^2}\left(x_p \sqrt{\gamma_{1,2}} \cos(q_n) + z \sin(q_n)\right)\right) \tag{15}$$

These accelerating solutions are traveling along a non-circular trajectory, bending to very large angles. This can be easily seen in $k$-space, where the transverse spatial frequencies vary while the beam is propagating in the $z$ direction. The change in the spatial frequencies can cause a propagating mode become evanescent while propagating in $z$. When this disappearance of modes occurs, the wavepacket is no longer shape-invariant. One of the most fascinating features is that the trajectory can even flip to the other direction – and accelerates towards the direction of the other lobes. The reason is that the metric changes in the $z$ direction, however after some z value this is no longer the direction normal to the wavefront, due to non-paraxial trajectory. This interesting feature could not be seen in the paraxial case. Naturally, this wavepackets is also constructed only from a discrete set of spatial frequencies: $q_n = \arccos\left(n/\sqrt{\gamma(q^2 \pm \kappa^2)}R_0\right)$. This has a major impact on the profile of the wavepackets – and it differs from that in flat space.

Finally, we return to the nonlinear case. For reasons of simplicity, we will deal here only with the paraxial regime. We seek a propagation-invariant solution of the paraxial nonlinear equation (Eq. (5)) traveling along a non-geodesic trajectory. Specifically, consider the Kerr effect, in which $V_{NL} = \kappa |\psi|^2 / \gamma$, where $\kappa$ is the effective nonlinear coefficient. We seek solutions (in the universal covering space) satisfying $|\psi(0, x)| = |\psi(z, x - f(z))|$ and obtain an equation for the amplitude of the beam $u(\tilde{x})$:

$$\tilde{u}_{\tilde{x}\tilde{x}} - \tilde{x}\tilde{u} + sign(\kappa)\tilde{u}^3 = 0 \tag{16}$$

where $\tilde{x} = \sqrt[3]{c}(x - f(z))$, $\tilde{u} = \kappa/c^{2/3} u$. The trajectory is the same as in the linear case (Eq. (7)). We solve Eq. (16) numerically for the focusing and defocusing case, ($\kappa > 0, \kappa < 0$ respectfully). The only free parameter in our solution is the initial conditions. To find the wavefunction, we assume that the nonlinear accelerating beam decays for $\tilde{x} \to \infty$, thus the nonlinear term in Eq. (16) is negligible for $\tilde{x} \to \infty$. Therefore we choose the initial condition to be $\tilde{u} = C \cdot Airy(\tilde{x})$ for $\tilde{x} \to \infty$. Typical shape-preserving solutions are shown in Fig. 3. These wavepackets are propagating in a self-similar fashion, similar to their linear counterparts. However, for the focusing case we find beams with narrower lobes than for the linear beam, and for the defocusing case we find beams with broader lobes. The solution for the focusing case exists for any $n_0 C^2 \kappa / k_0^2 > 0$, whereas for the defocusing case the solution exists only for a finite range of $-2.5 < n_0 \kappa C^2 / k_0^2 < 0$. (in accordance to [26]). Next, we check the stability of our solutions and find the solution for the defocusing case to be stable under random "white" noise, whereas the self-focusing solutions become unstable after some propagation distance. The most interesting feature is that the stability of the beam depends on the curvature: by changing the parameters of the surface we can make the beam stable for considerably larger distances (possibly even indefinite), as shown in Fig 3. This suggest on option for stabilizing nonlinear accelerating beams using the curvature of space, this is directly related to the instability of solitons in negatively curved space [44]. To augment this nonlinear section, we note that other saturable nonlinearities can be handled in a similar fashion, as was done in [26] for flat space.

To summarize, we have found linear and nonlinear, paraxial and non-paraxial, spatially-accelerating wavepackets in curved space, thereby introducing the concept of accelerating beams to curved space geometry. This work raises many further interesting ideas. The relation of this work to General Relativity opens up many ideas for future exploration. For example, the current work shows that wavepackets in curved space can be controlled by specifically designing their input wavefront. In principle, this means that one can predesign a wavefront that would be able to overcome (compensate for) effects of Gravity. Indeed, we are currently working on the nonlinear version of this idea, where the accelerating wavepacket is what causes the effective curving in space, in an optically nonlinear medium.

This work was supported by the ICORE Excellence Center "Circle of Light", by an Advanced Grant from the European Research Council, by the Israel Ministry of Science and Technology and by the Israel Science Foundation.

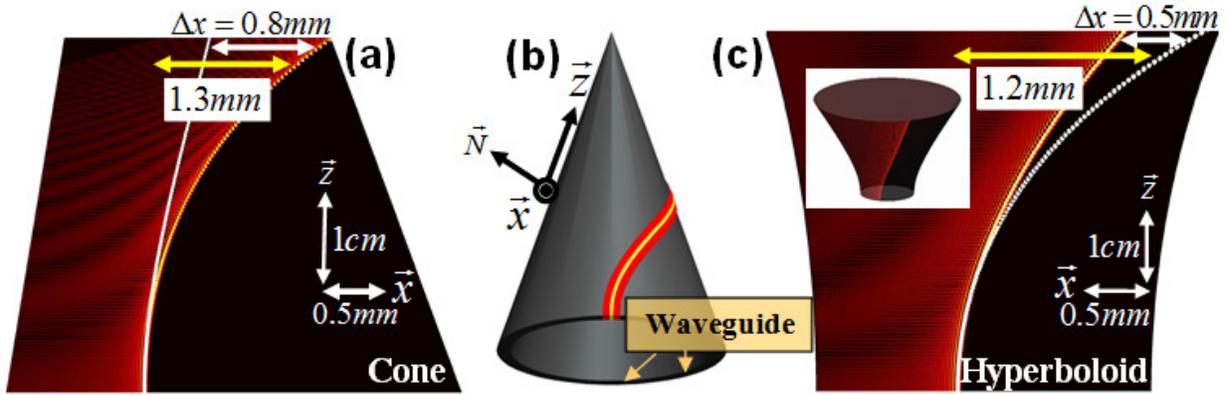

**FIG. 1.** (b) Sketch of a surface of revolution (cone). The EM field is restricted to propagate within the surface area by a thin waveguide layer. (a)-(c) The evolution of the envelope of an accelerating beam $(\psi)$ on the surface area of a cone (a), and of an hyperboloid (both with $10\ cm$ height, $3\ mm$ base radius and a propagation constant of $1.2e7 m^{-1}$). (c) Dashed white line displays the propagation of the same beam in flat space, projected on the surface of revolution. The beam aperture is $9\ mm$ (a) and $6\ mm$ (c), and a main lobe of width of $30\ \mu m$ (a) $33\ \mu m$ (c).

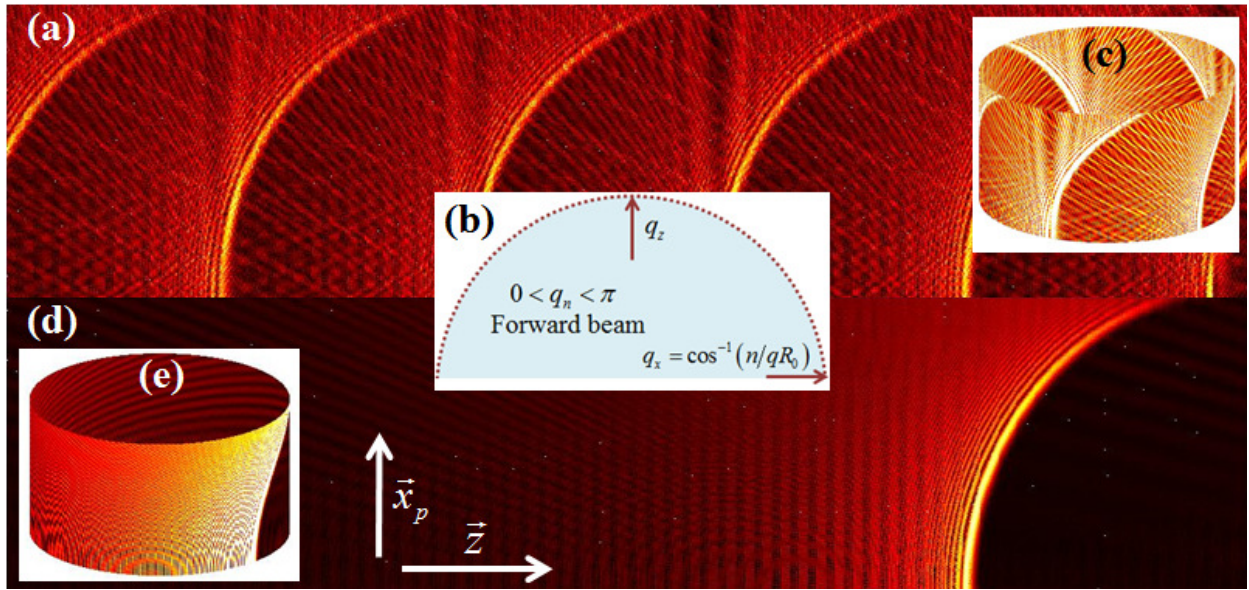

**FIG. 2.** Periodic accelerating beams constructed from discrete spatial frequencies, as they propagate on various cylinders. This wavepackets travels along circular trajectories while bending to very large non-paraxial angles. (a) The wavepacket constitutes of several parallel beams that depend on the initial choice of $D_n$; only every fifth spectral function is populated, each with $D_{5n}=1$. (b) Schematic illustration of the periodic accelerating beam in k-space. The beam in constructed from discrete spatial frequencies that reside in a half-circle in k-space. This is a superposition of only forward propagating waves. (d) Periodic accelerating beam on a cylinder, displaying a single intense main lobe $(D_n=1)$. (c),(e) The beams from (a) and (d) propagating on a surface of a cylinder.

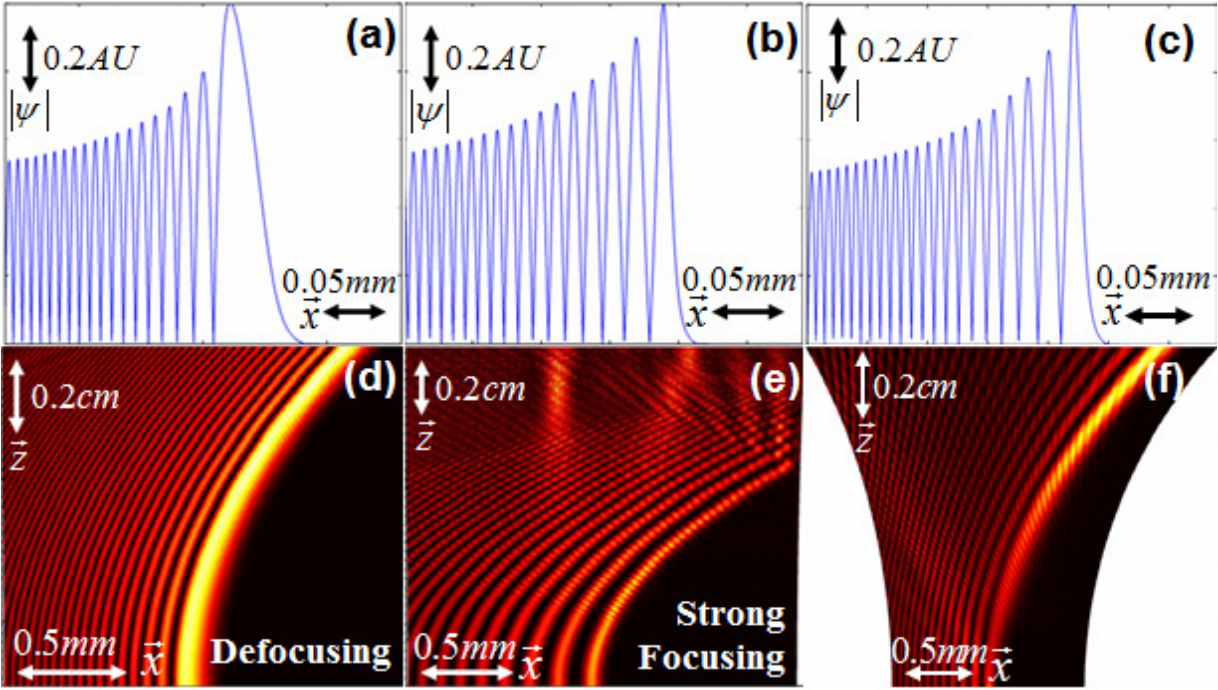

**FIG. 3.** (a)-(c) Profile of a nonlinear accelerating wavepackets in curved space under defocusing (a) and focusing (b)-(c) Kerr nonlinearity. The profile differs from the nonlinear accelerating beam: for the defocusing case the lobes are wider than the linear Airy beam whereas for the focusing case the lobes are much thinner. (d)-(f) Evolution of the nonlinear wavepackets (of (a)-(c)) on the surface area of an hyperboloid with random noise (of 2%). Defocusing nonlinearity supports stable propagation (d), whereas for strong focusing (e) the beam become unstable to noise. (f) Evolution of the beam of (e) under different surface parameters can make the beam stable.